# A Study and Preliminary Model of Cross-Domain Influences on Creativity

**Liane Gabora (liane.gabora@ubc.ca)**
Department of Psychology, University of British Columbia
Okanagan Campus, Arts Building, 1147 Research Way, Kelowna BC, V1V 1V7, CANADA

**Nicole Carbert (nicole.carbert@alumni.ubc.ca)**
School of Population and Public Health, Faculty of Medicine, University of British Columbia
Vancouver BC, V6H 3V4, CANADA

**Abstract**

This paper takes a two-pronged approach to investigate the phenomenon of cross-domain influence on creativity. We present a study in which creative individuals were asked to list influences on their creative work. More than half the listed influences were unrelated to their creative domain, thus demonstrating empirically that cross-domain influence is widespread. We then present a preliminary model of exaptation, a form of cross-domain influence on creativity in which a different context suggests a new use for an existing item, using an example from the study.

## Introduction

The study of cross-domain thinking in cognitive science has focused largely on analogy and metaphor, but the phenonmenon extends further. One indication of this is a tradition in the arts, referred to as *ekphrastic expression,* of interpreting art from one medium (*e.g*., acrylic painting) into another (*e.g*., charcoal sketch). The goal of ekphrastic expression is to capture, and thereby become intimate with, the underlying form or essence of a work by translating it from one medium into another, and have a more direct impact on an audience. A related phenomenon is that of *cross-media style*, wherein the same style is demonstrated by works of art in multiple media. The term rococo, for example, is given to a style of painting, sculpture, literature, and music of the 18$^{th}$ Century. It is thought that works in a particular style suggest underlying abstract archetypal forms or potentialities to the artistic mind that compel the exploration of different manifestations (Burke, 1957).

The phenomena of ekphrastic expression and cross-media style are consistent with evidence that creative works in different media may be similar in terms of psychophysical, collative, and ecological properties (Hasenfus, 1978). Aesthetic perceptions stimulated by creative works may generate physiological, emotional, cognitive, and/or behavioral, responses that are amenable to re-expression in other forms. This may be due to regularities in the choice of elements (*e.g.*, shapes, colors, or words) and/or how they are used (*e.g*., in a chaotic or orderly manner) (Berlyne, 1971). It has been shown that there are non-arbitrary mapping between properties of vision and sound (Griscom & Palmer, 2012; Melara, 1989; Melara, & Mark, 1990; Palmer, Schloss, Xu, & Prado-Leon, 2013; Ward, Huckstep, & Tsakanikos, 2006). For example, the processing of visual features, such as lightness and spatial frequency, can be affected by auditory features such as pitch and timbre (Marks, 1974, 1975, 1987).

In a study of cross-domain creativity that aimed to move beyond single-dimensional mappings, composers were asked to write music inspired by four simple line-drawn shapes: a square, a lightning bolt, a curvy shape, and a jagged shape (Willmann, 1944). Music inspired by the same shape was more similar than music inspired by another shape with respect to tempo, melodic pattern, mood, and other characteristics, and listeners could match above chance the music to the shape that inspired it. However, the impoverished nature of the stimuli undoubtedly limited the scope for creative expression. Another study aimed at investigating whether the rich emotionality of genuinely creative works could be translated to, and recognized in another domain. It demonstrated that when pieces of music were re-interpreted as paintings, naïve participants were able to correctly identify at significantly above chance which piece of music inspired which painting (Ranjan, Gabora, & O'Connor, 2014; Ranjan, 2014). Although the medium of expression is different, something of its essence remains sufficiently intact for an observer to detect a resemblance between the new work and the source that inspired it. This result lent empirical support to the largely anecdotal body of evidence that cross-domain influence is a genuine phenomenon, and suggested that, at their core, creative ideas are less domain-dependent than they are generally assumed to be. It did not, however, provide evidence that the phenomenon extends beyond the artificial conditions of such a study, nor did it give an indication of how prevalent it is.

## A Study of Cross-Domain Influence on Creative Innovation

The goal of the present study was to provide a preliminary assessment of the extent to which creative individuals are influenced by stimuli and experiences that are directly related, indirectly related, and unrelated to their domain of creative expression.

## Method

An internet search was conducted to locate individuals who are creative in any domain. They were recruited by email and invited to participate in the study on a voluntary basis. The message provided a link to an online questionnaire that was hosted by SurveyMonkey. There was no remuneration for participation. The questionnaire asked their gender, age, and occupation, as well as the following questions:

1. What is the general category for the creative work for which you are most known (e.g., art, music, drama, science)?
2. What is the subcategory for the creative work for which you are most known (e.g., painting, piano composition, biochemistry)?
3. Please describe your creative outputs.
4. Please describe as best you can your creative process.
5. Describe all elements that have inspired your work (natural or artificial, or it may be a particular event or situation, or something not in the concrete environment, that is, something abstract that you have been thinking about), and with each item, if possible, put as much identifying information as you can about the item it inspired (e.g., my Sunlight Sonata in B Flat composed in 2012 was inspired by going skiing in the alps with my sister who had just recovered from pneumonia). Do this for as many of your creative works as you can, itemizing them as (a), (b), (c), and so forth. Provide as much detail as possible.

The first three questions were used to categorize the creators into the following primary creative domains: art, music, and writing. Artists were further categorized into secondary domains: painting, drawing, photography, and sculpture. Question four was not used in this analysis. Responses to question five were divided into four categories: cross-domain, within-domain narrow, within-domain broad, and uncertain. The last category was used when not enough information had been provided to make a distinction, i.e., the influence could have been either within-domain or cross-domain.

## Results

66 individuals (50 females and 16 males) completed the questionnaire. They provided a total of 65 influences (i.e., almost one influence listed per participant). Examples of each category of influence are provided in Table 1. The total number of influences in each category is provided in Table 2. For some domain-category combinations there were no examples given by any participant.

The frequency of cross-domain influences (47%) was greater than that of within-domain influences (27%), and this was the case even when broad as well as narrowly construed within-domain influences were considered (35%).

Table 1: Examples from the data of each of the four categories of influence. Top: narrow within-domain (WD-n) and broad within-domain (WD-b) influences. Bottom: cross-domain influences (CD) and influences categorized as "uncertain" (U). A dash indicates that no examples of that category were present in the data.

| **Creator** | **WD-n** | **WD-b** |
|---|---|---|
| Artist - Painting | Galleries | Spirograph |
| Artist - Drawing | Political cartoonists | – |
| Artist - Photography | – | Books and lectures on subject of "understanding pictures" |
| Artist -Sculpture | – | Architectural elements |
| Musician | Band musician collaboration | – |
| Writer | Conferences | – |

| **Creator** | **CD** | **U** |
|---|---|---|
| Artist - Painting | Global warming | Opposites |
| Artist - Drawing | Comedy | Circular intellect |
| Artist - Photography | Meditation | – |
| Artist -Sculpture | Computer programming | World |
| Musician | Literature | Creativity seminar |
| Writer | Nature | Retreats |

Table 2: Number of participants in each creative domain (N), and the raw number (r) and percentage (%) of influences that were cross-domain (CD), within-domain: narrow (WD-n), within-domain: broad (WD-b, and uncertain (U). Percentages are in brackets. A dash indicates that no examples were present in the data.

| Creative Domain | N | CD r | CD (%) | WD-n r | WD-n (%) | WD-b r | WD-b (%) | U r | U (%) |
|---|---|---|---|---|---|---|---|---|---|
| Painting | 44 | 21 | (48) | 12 | (27) | 4 | (9) | 6 | (14) |
| Drawing | 8 | 2 | (25) | 2 | (25) | – | – | 3 | (38) |
| Photography | 4 | 2 | (50) | – | – | 1 | (25) | – | – |
| Sculpture | 5 | 3 | (60) | – | – | – | – | 1 | (20) |
| Music | 3 | 1 | (33) | 2 | (68) | – | – | 1 | (33) |
| Writing | 2 | 2 | (100) | 1 | (50) | – | – | 1 | (50) |
| TOTAL | 66 | 31 | **(47)** | 17 | **(27)** | 5 | **(8)** | 12 | **(18)** |

## Discussion

These results demonstrate that even if individuals primarily express their creativity in a single domain, they are often employing cross-domain thinking when they create. The study enriches our understanding of how the creative process works by adding to a growing body of evidence that creativity is not just a matter of acquiring domain-specific expertise. Two limitations of this study are the small sample size and the focus on artistic creativity. We are currently attempting to obtain more data, focusing our efforts on individuals who are scientifically and technically creative.

## A Quantum Model of Cross-Domain Influence on Creative Innovation

An interesting form of cross-domain influence is *exaptation,* wherein a trait that originally came about to solve one problem is co-opted for another use. The concept of exaptation comes from biology but has been shown to play a pivotal role in economics (Dew, Sarasvathy, & Ventakaraman, 2004). A preliminary attempt has been made to develop a mathematical model of exaptation that can be applied across disciplines (Gabora, Scott, & Kauffman, 2013). Here we use it to model an example provided by an artist in the above study who used scraps of tissue paper to make clouds in a mixed media artwork.

The model we use is a generalization of the formalism of quantum mechanics adapted for application in a psychological context (Aerts & Gabora, 2005; Aerts, Gabora, & Sozzo, 2013; Busemeyer & Bruza, 2012; Pothos & Busemeyer, 2013).[1] Quantum probability models in psychology have been compared side-by-side with classical models (Busemeyer, Pothos, Franco, & Trueblood, 2011). According to classic probability, all events are subsets of a common sample space; that is, they are based on a common set of elementary events. An important advantage of a quantum model over a classical model such as a Bayesian one is that it uses variables and spaces that are defined specifically with respect to a particular context, which is necessary to capture certain aspects of how concepts behave (Aerts & Gabora, 2005; Gabora & Aerts, 2002; Kitto, Ramm, Sitbon, & Bruza, 2011). The state $|\psi\rangle$ of an entity is written as a linear superposition of a set of basis states $\{|\phi_i\rangle\}$ of a Hilbert space $\mathcal{H}$, which is a complex vector space with an inner product.[2] Another advantage of a quantum model over a classical one is that it uses *amplitudes*, which though directly related to probabilities, can exhibit interference, superposition, and entanglement, which are also needed to capture certain aspects of how concepts behave (Aerts, 2009; Aerts, Broekaert, Gabora, & Veloz, 2012; Aerts, Gabora, & Sozzo, 2013; Aerts & Sozzo, 2011; Bruza, Kitto, Ramm, & Sitbon, 2011). The amplitude term, denoted $a_i$, is a complex number that represents the contribution of a component state $|\phi_i\rangle$ to the state $|\psi\rangle$. Hence $|\psi\rangle = \Sigma_i a_i |\phi_i\rangle$. The square of the absolute value of the amplitude equals the probability that the state changes to that particular component basis state. A non-unitary change of state is called *collapse*. The square of the absolute value of the amplitude equals the probability that the state changes to that particular component basis state upon measurement. The choice of basis states is determined by the observable $O$ to be measured, and its possible outcomes $o_i$. The basis states corresponding to an observable are referred to as *eigenstates*. Observables are represented by self-adjoint operators on the Hilbert space. The lowest energy state is referred to as the ground state. Upon measurement, the state of the entity collapses from its current state (possibly the ground state) and is projected onto one of the eigenstates.

Now consider two entities $A$ and $B$ with Hilbert spaces $\mathcal{H}_A$ and $\mathcal{H}_B$. We denote amplitudes associated with the first and second as $a_i$ and $b_j$ respectively. The Hilbert space of the composite of these entities is given by the tensor product $\mathcal{H}_A \otimes \mathcal{H}_B$. We may define a basis $|e\rangle_i$ for $\mathcal{H}_A$ and a basis $|f\rangle_j$ for $\mathcal{H}_B$. The most general state in $\mathcal{H}_A \otimes \mathcal{H}_B$ has the form

$$|\psi\rangle_{AB} = \Sigma_{i,j}\, c_{ij}\, |e\rangle_i \otimes |f\rangle_j \qquad (1)$$

where $c_{ij}$ is the amplitude corresponding to the composite entity.

The phenomenon of *entanglement* was conceived to deal with situations of non-separability where different entities form a composite entity. The state $|\psi\rangle_{AB}$ is separable if for the amplitudes $c_{ij}$ amplitudes $a_i$ and $b_j$ can be found such that $c_{ij} = a_i\, b_j$. It is inseparable, and therefore an entangled state, if this is not possible, hence if the amplitudes describing the state of the composite entity are not of a product form.[3] Entangled states are non-compositional because they may exhibit emergent properties not inherited from their constituent components.

It is not exactly the quantum formalism that we use but a generalization (sometimes called "quantum-like") that is adapted for application to concepts. In quantum inspired models of concepts, a context plays the role of a measurement. A set of basis states related to a context represents instances of a concept. A context can exert either a deterministic or probabilistic influence on the state of a concept. If there is no uncertainty or choice involved then the change of state is deterministic and this is represented by a linear operator, which may be a unitary operator, a projection operator, or an operator of a more general nature, depending on the type of contextual influence. If there is uncertainty or choice involved then the change of state is probabilistic. Different possible outcomes can occur, each with a certain probability, and the effect of context is represented by a self-adjoint operator.

[1] This approach is unrelated to quantum models of consciousness (Hammeroff, 1998) or memory (Pribram, 1993), and makes no assumption that phenomena at the quantum level affect the brain; it draws solely on abstract formal structures that, as it happens, found their first application in quantum mechanics.

[2] It is slightly more complex but more accurate to define a Hilbert space as a real or complex inner product space that is also a complete metric space with respect to the distance function induced by the inner product. The inner product allows one to define the length of a vector and the angle between two vectors, as well as orthogonality between vectors (zero inner product).

[3] In some applications the procedure for describing entanglement is more complicated than what is described here. For example, it has been argued that quantum field theory, which uses Fock space to describe multiple entities, gives an internal structure that is superior to the tensor product for modeling concept combination (Aerts, 2009). Fock space is the direct sum of tensor products of Hilbert spaces, so it is also a Hilbert space.

In one generalized quantum formalism, namely the State Context Property (SCOP) theory of concepts, a concept is defined in terms of (1) its set of states $\Sigma$ (including both exemplars and ground states changed under the influence of a context), a set $\mathcal{L}$ of relevant properties or features, (3) a set $\mathcal{M}$ of contexts in which the concept may be relevant, (4) a function $\nu$ that gives the applicability or *weight* of a certain property for a particular state and context, and (5) a function $\mu$ that gives the probability of transition from one state to another under the influence of a particular context. We might represent the state of a chair by a vector $|p\rangle$ of length equal to 1 in a complex Hilbert space $\mathcal{H}$. From a different context $|p\rangle$ could actualize as another state. For example, in the context office it may actualize as OFFICE CHAIR, while in the context kitchen it may actualize as KITCHEN CHAIR. These are deterministic changes of state.

More interesting is a probabilistic change of state in which there are two or more possible outcomes. For example, consider the reconceptualization of scraps of tissue paper described by one of our respondents to question five. The concept TISSUE PAPER could change probabilistically from $|p\rangle$ to one of two states: $|u\rangle$, the state in which it is viewed as *useful* (*e.g.*, invent a new use for it), and $|w\rangle$, the state in which it is viewed as *waste*. The state of TISSUE PAPER prior to being conceived of as useful or waste is modeled as a superposition of these two possibilities. The vectors $|u\rangle$ and $|w\rangle$ form the basis of a complex Hilbert space. Thus state $|p\rangle$ of TISSUE PAPER can be written as a superposition of $|u\rangle$ and $|w\rangle$, *i.e.*,

$$|p\rangle = a_0|u\rangle + a_1|w\rangle \quad (2)$$

where $a_0$ and $a_1$ are complex numbers that give the amplitudes of $|u\rangle$ and $|w\rangle$ respectively. More concretely, the probability that $|p\rangle$ is viewed as useful equals $|a_0|^2$, the square of the absolute value of $a_0$. The probability it is viewed as waste equals $|a_1|^2$, the square of the absolute value of $a_1$. If it were to be decided that TISSUE PAPER is useful or waste the state would change probabilistically from $|p\rangle$ to $|u\rangle$ or $|w\rangle$. The states $|u\rangle$ and $|w\rangle$ are thus eigenstates of TISSUE PAPER in a default, generic context. In different individuals $a_0$ and $a_1$ may have different values (as epitomized in the saying, "one person's trash is another person's treasure"). TISSUE PAPER may also be conceived differently by the same person in different modes of thought. Divergent thinking may facilitate the process of finding a new use for TISSUE PAPER, as recounted in the following excerpt from an artist's description of the creative process that gave rise to a mixed media artwork titled 'Encouragement':

> I was going to use paint… when the painting insisted I use prepared tissue paper instead. We had an argument. … The thought persisted and having had some previous experience arguing with a painting I decided to go with the suggestion... I must say I am very pleased with the results.

When shreds of tissue paper are considered in the context of the artwork titled Encouragement, denoted $e$, this changes the likelihood of the tissue paper being conceived of as useful or waste. Recall that states are represented by unit vectors, and all vectors of a decomposition such as $|u\rangle$ and $|w\rangle$ have unit length, are mutually orthogonal, and generate the whole vector space; thus $|a_0|^2 + |a_1|^2 = 1$. This means that the change in the probability that TISSUE PAPER is viewed as useful if one considers it from a new context can be modeled using a Pythagorean argument, as in Figure 1.

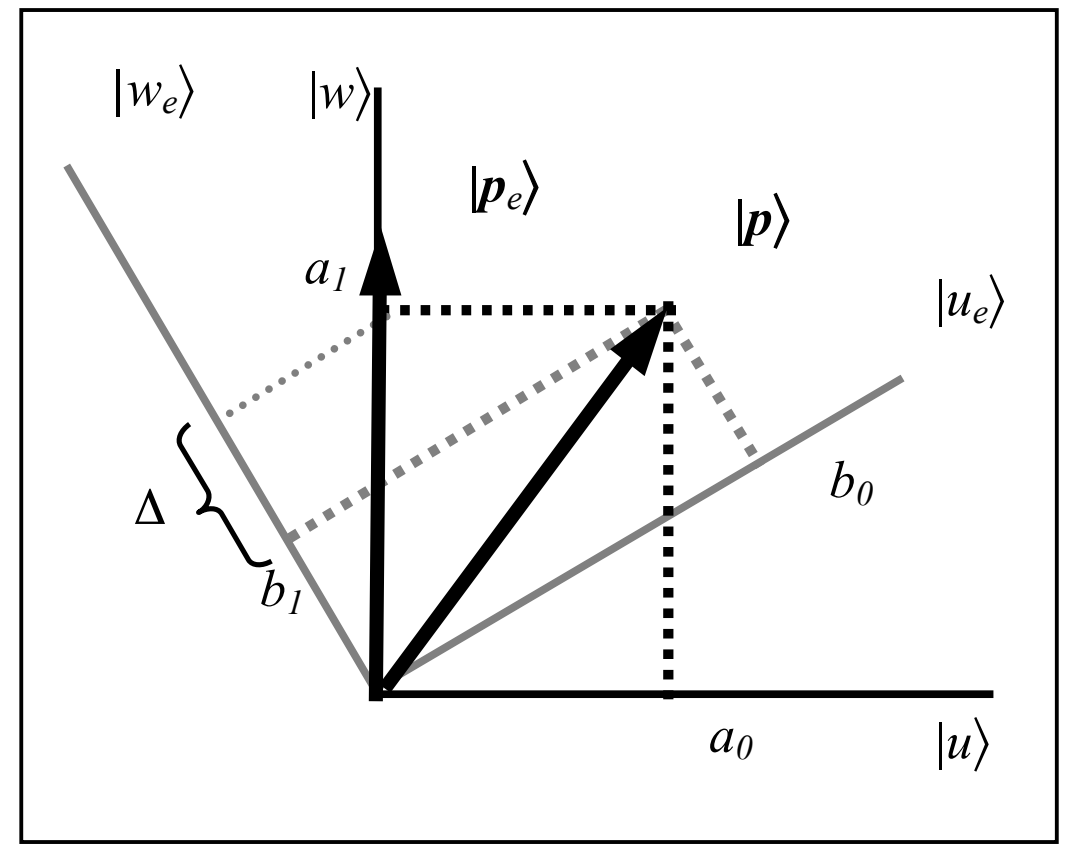

Figure 4: Graphical depiction of a vector $|p\rangle$ representing the state of TISSUE PAPER. In the default context it is likely to collapse to projection vector $|w\rangle$ which represents that it is waste. This can be seen by the fact that subspace $a_0$ is smaller than subspace $a_1$. Thinking more creatively one might consider using TISSUE PAPER in an artwork. Thus in the context of Encouragement (shown in gray), the state of TISSUE PAPER is likely to collapse to the orthogonal projection vector $|u\rangle$ which represents that it is useful, as shown by the fact that $b_0$ is larger than $b_1$. Also shown is the projection vector after renormalization (the vertical arrow).

In the context of the artwork titled Encouragement, creative states of TISSUE PAPER are to use it to depict items in a landscape that, to the artist, help convey a feeling of encouragement, such as perhaps clouds or houses. Let us show that the formalism is capable of incorporating these possibilities as states of the concept TISSUE PAPER. This will not be possible in the Hilbert space formed by the two states $|u\rangle$ and $|w\rangle$ because it has only two dimensions. The restructured conception of TISSUE PAPER in the context Encouragement, denoted $|p_e\rangle$, is given by

$$|p_e\rangle = a_3 \mathrm{P}u_e\, |p_e\rangle / \|\, \mathrm{P}u_e\, |p_e\rangle \| + a_4 |w_e\rangle \quad (3)$$

where $\mathrm{P}u_e$ is an orthogonal projection operator. We substitute in the mathematical formalism of Hilbert space for the unit vector whenever what physicists call ‘degeneration’ is involved, meaning that several orthogonal states can give rise to the same property, here the property *useful.* Note that $\|Pu_e\, |p_e\rangle\|$ is the length of $Pu_e\, |p_e\rangle$. We need to divide the vector $Pu_e\, |p_e\rangle$ by $\|Pu_e\, |p_e\rangle\|$ for it to become a unit vector, and hence represent a state. Let us specify these states of usefulness to make the mathematical description complete. Since we want to consider creative useful states, specifically CLOUDS and HOUSES, we introduce the states $|c\rangle$ and $|h\rangle$ respectively. In the context of the artwork titled Encouragement, they are denoted $|c_e\rangle$ and $|h_e\rangle$. $|t_e\rangle$ denotes the possibility that, even in the context Encouragement, TISSUE PAPER is viewed as useful just as it is (without using it to represent anything). $|t_e\, h_e\rangle$ and $|t_e\, c_e\rangle$ denote the possibility that in the context Encouragement TISSUE PAPER is used to depict houses and clouds, respectively. We write the projector as the sum of the partial projectors on the states. Hence we have

$$Pu_m = |t_e\rangle\langle t_e| + |t_e\, h_e\rangle\langle t_e\, h_e| + |t_e\, c_e\rangle\langle t_e\, c_e| \qquad (4)$$

where $|t_e\rangle\langle t_e|$, $|t_e\ h_e\rangle\langle t_e\ h_e|$ and $|t_m\ c_m\rangle\langle t_e\ c_e|$ are the one dimensional orthogonal projection operators on the vectors $|t_e\rangle$, $|t_e\, h_e\rangle$ and $|t_e\, c_e\rangle$ respectively. By considering TISSUE PAPER in different contexts, the perceived probability that they are useful has increased, *i.e.*, $|a_3|^2 > |a_0|^2$ because the state $Pu_e\ |p_e\rangle/\|\ Pu_e\ |p_e\rangle\|$ incorporates possibilities of representing houses or clouds.

TISSUE PAPER has a set $\mathcal{L}$ of features that includes, for example, ‘opaque’, denoted $f_1$. The states HOUSES and CLOUDS possess features as well. For example, the set of features of HOUSES includes ‘rigid’, denoted $f_2$.

Tissue paper is not rigid, so $\nu(p, f_2) << \nu(h, f_2)$. Writing the unit vector $Pu_e\, |p_e\rangle/\|\ Pu_e\, |p_e\rangle\|$ again as a superposition of vectors $|t_e\rangle$, $|t_e\, h_e\rangle$ and $|t_e\, c_e\rangle$ we have:

$$Pu_e\, |p_e\rangle/\|\ Pu_e\, |p_e\rangle\| = a_5|t_e\rangle + a_6|t_e\, h_e\rangle + a_7|t_e\, c_e\rangle \qquad (5)$$

Because tissue paper is not rigid, $|a_6|^2$ is small.

However, CLOUDS do not possess the feature of rigidity, $f_2$. Moreover CLOUDS *can* possess $f_1$, opacity. Therefore, $\nu(p, f_1) \approx \nu(c, f_1)$ and $|a_7|^2$ is large. This means that $\mu(c, e, p) >> \mu(h, e, p)$. Thus, in the context of the artwork titled Encouragement, the concept TISSUE PAPER has a high probability of collapsing to TISSUE PAPER CLOUDS.

We can model the emergence of new properties using the notion of entanglement. Although the state TISSUE PAPER CLOUDS was modeled by $t_e\, c_e\rangle$ as one of the sub-states of TISSUE PAPER, the quantum formalism can also be used to derive this state as a combined state of TISSUE PAPER and CLOUD. It has been shown experimentally that such a combined state is in general not a product state but an entangled state (Aerts, Gabora & Sozzo, 2013, Aerts & Sozzo, 2011). Thus, following the mathematical formalism of quantum theory, TISSUE PAPER CLOUDS may actualize new properties that are not normally properties of tissue paper or clouds, such as “sparkle-covered” (using sparkles to convey the look of sun shining through).

### Discussion

This example shows that it is possible to model the creative restructuring of a concept (e.g., TISSUE PAPER) when it is considered from a new perspective (e.g., using it to depict a scene that expresses “encouragement”). This example of cross-domain thinking is so subtle that it had not even been categorized as such in the study, showing that cross-domain thinking is even more widespread than what the study revealed. The example is simple; there is much work to be done to model the complex ways in which new situations influence how one “puts a new spin” on a product or idea. Nonetheless, the approach provides a formal model of what Rothberg (2015) calls Janusian thinking. It adds to a growing literature that uses quantum methods to model with experimental data on how people use concepts (e.g., Aerts 2009; Aerts, Aerts, & Gabora, 2009; Aerts, Gabora, & Sozzo, 2013), which has been expanded to incorporate larger conceptual structures (Gabora & Aerts, 2009), as well as how the same concepts are conceived of differently in divergent versus convergent modes of thought (Veloz, Gabora, Eyjolfson, & Aerts, 2011). In the quantum representation, probability is treated as arising not from a lack of information per se, but from the limitations of any context (even a ‘default’ context). Note that it is not appropriate to describe the artist’s cognitive state prior to realizing that tissue paper could be used to make clouds as a collection of separate, discrete possibilities, some of which incorporate this idea. The formalism allows us to capture the potentiality inherent in cognitive states of this type.

## General Discussion and Conclusions

This paper outlined a multi-faceted initial attempt to investigate the phenomenon of cross-domain influence on creativity. First, we carried out a study that showed empirically that the phenomenon is indeed widespread. Second, using an example taken from the study, we made a preliminary model of a simple form of cross-domain creative influence employing the concept of exaptation wherein an item is seen to possess new affordances when it is imported to a new context. Thus we provided proof of concept that it is possible to begin to model cross-domain creative thought processes.

It is hoped that these complimentary directions pave the way to a deeper understanding of how our richly complex world influences the process by which that rich complexity is amplified. We are perhaps just beginning to understand the myriad channels of influence that result in the creation of an artistic masterpieces or technological feat.

## Acknowledgments

We are grateful for funding to the first author from the Natural Sciences and Engineering Research Council of Canada.